\documentclass[iop]{emulateapj}
\usepackage{graphicx}
\usepackage{mediabb}

\newcommand{\swift}{\textit{Swift}}

\newcommand{\fermi}{\textit{Fermi}-LAT}

\newcommand{\src}{3C~120}

\slugcomment{Accepted by ApJL on 7 January, 2015.}

\begin{document}

\title{Six years of Fermi-LAT and Multi-wavelength Monitoring of the Broad-Line Radio Galaxy 3C~120: Jet Dissipation at Sub-parsec scales from the Central Engine}

\author{Y.~T.~Tanaka\altaffilmark{1}, A.~Doi\altaffilmark{2}, Y.~Inoue\altaffilmark{2}, C.~C.~Cheung\altaffilmark{3}, L.~Stawarz\altaffilmark{2, 4}, Y.~Fukazawa\altaffilmark{5}, M.~A.~Gurwell\altaffilmark{6}, M.~Tahara\altaffilmark{7}, J.~Kataoka\altaffilmark{7}, and R.~Itoh\altaffilmark{5}}

\email{ytanaka@hep01.hepl.hiroshima-u.ac.jp}

\altaffiltext{1}{Hiroshima Astrophysical Science Center, Hiroshima University, 1-3-1 Kagamiyama, Higashi-Hiroshima 739-8526, Japan}
\altaffiltext{2}{Institute of Space and Astronautical Science, JAXA, 3-1-1 Yoshinodai, Chuo-ku, Sagamihara, Kanagawa 252-5210, Japan}
\altaffiltext{3}{Space Science Division, Naval Research Laboratory, Washington, DC 20375-5352, USA}
\altaffiltext{4}{Astronomical Observatory, Jagiellonian University, ul. Orla 171, 30-244 Krak\'ow, Poland}
\altaffiltext{5}{Department of Physical Sciences, Hiroshima University, Higashi-Hiroshima, Hiroshima 739-8526, Japan}
\altaffiltext{6}{Harvard-Smithsonian Center for Astrophysics, Cambridge, MA 02138, USA}
\altaffiltext{7}{Research Institute for Science and Engineering, Waseda University, Tokyo 169-8555, Japan}

\begin{abstract}
We present multi-wavelength monitoring results for the broad-line radio galaxy \src\ in the MeV/GeV, sub-millimeter, and 43~GHz bands over six years. Over the past two years, \fermi\ sporadically detected \src\ with high significance and the 230~GHz data also suggest an enhanced activity of the source. After the MeV/GeV detection from \src\ in MJD 56240--56300, 43 GHz VLBA monitoring revealed a brightening of the radio core, followed by the ejection of a superluminal knot. Since we observed the $\gamma$-ray and VLBA phenomena in temporal proximity to each other, it is naturally assumed that they are physically connected. This assumption was further supported by the subsequent observation that the 43~GHz core brightened again after a $\gamma$-ray flare occurred around MJD 56560. We can then infer that the MeV/GeV emission took place inside an unresolved 43~GHz core of \src\ and that the jet dissipation occurred at sub-parsec distances from the central black hole, if we take the distance of the 43~GHz core from the central black hole as $\sim 0.5$~pc, as previously estimated from the time lag between X-ray dips and knot ejections \citep{Marscher02,Chatterjee09}. Based on our constraints on the relative locations of the emission regions and energetic arguments, we conclude that the $\gamma$ rays are more favorably produced via the synchrotron self-Compton process, rather than inverse Compton scattering of external photons coming from the broad line region or hot dusty torus. We also derived the electron distribution and magnetic field by modeling the simultaneous broadband spectrum.
\end{abstract}


\clearpage

\section{Introduction}
\label{sec-intro}

The Fanaroff-Riley Class I radio galaxy \citep{Walker87} \src\ ($z=0.033$) is optically classified as a broad-line radio galaxy (BLRG). Its broadband spectrum is dominated by the non-thermal jet emission along with a thermal disk component, hence this BLRG is an ideal source to study the jet-disk connection. Indeed, \citet{Marscher02} observed X-ray dips in \src\ followed by radio knot ejections through Very Long Baseline Array (VLBA) monitoring. This was naturally interpreted as disk material suddenly falling onto its central black hole (BH) with some fraction ejected as a jet \citep[see also][]{Chatterjee09}. This jet-cycle paradigm in \src\ has been confirmed independently by multi-epoch X-ray spectroscopy combined with simultaneous UV and radio observations \citep{Lohfink13}.

The initial \textit{Fermi} Large Area Telescope \citep[LAT;][]{Atwood09} detection of \src\ was reported by \citet{Abdo10MAGN}. \citet{Kataoka11} argued that the $\gamma$-ray emissions of LAT-detected BLRGs (\src, 3C~111 and Pictor~A) are most likely produced in the inner nuclear jets rather than large-scale jet structures, because these sources contain brighter radio cores compared to other BLRGs not detected by the LAT \citep[see also][for 3C~111]{Grandi12}. A major question is where the $\gamma$-ray emission is located relative to the central BH. It has been assumed that a measurement of the variability timescale directly constrains the location of the emission if the jet is conical and the emission region roughly corresponds to the cross-sectional radius of the jet. However, the detection of rapid day-scale TeV flaring observed in 2005 from M87 \citep{M87_10years_12} was determined to arise from the pc-scale knot known as HST-1 located at a projected distance of 65~pc from the central BH rather than the VLBA core \citep{Cheung07}. The recent detection of rapidly variable (doubling time of 10 minutes) sub-TeV emission from one of few TeV-emitting flat spectrum radio quasars 4C +21.35 also indicates that a very compact emission region of the order of 10$^{14}$ cm is located at several pc from the BH to avoid $\gamma\gamma$ annihilation with broad line region photons \citep{Aleksic11,Tanaka11,Tavecchio11}. Therefore, the variability timescale does not necessarily directly relate to the jet dissipation location, and hence a multi-wavelength approach is crucial to determining the MeV/GeV $\gamma$-ray production site.

How powerful relativistic jets are launched from a supermassive BH is also a key issue regarding Active Galactic Nuclei (AGNs). A likely jet launching mechanism was presented by \citet{Blandford77}, who considered that BH rotational energy is extracted magnetically and converted to jet power. Magneto-centrifugal force is also thought to drive the jet in part \citep{BP82}. Recent theoretical studies have shown that rapidly rotating BHs in magnetically arrested accretion can efficiently generate relativistic jets \citep[e.g.,][]{Tche11, Mckinney12, Sikora13}.

In this Letter we present six years of \fermi\ monitoring results of \src, together with Submillimeter Array (SMA) 230~GHz and VLBA 43~GHz light curves. We describe the observations and data reduction in \S\ref{sec-data}. Based on the multi-wavelength results presented in \S\ref{sec-results}, we discuss the location of the jet dissipation region as well as the MeV/GeV $\gamma$-ray emission mechanism in \S\ref{sec-dis}. The relation between the power in the jet and the disk is presented in \S\ref{sec-dis}.

\section{Observations}
\label{sec-data}

\subsection{\fermi}
\label{sec-LAT}

Six years of \fermi\ \texttt{P7REP} data from 2008 August 4 to 2014 August 4 were analyzed using the Fermi ScienceTools v9r33p3. We selected \texttt{SOURCE} class events using the instrument response functions \texttt{P7REP\char`_SOURCE\char`_V15} within a 10$^{\circ}$ region of interest (ROI) centered at the radio position of \src\ (${\rm R.A.}=68.296^{\circ}$, ${\rm Decl.}=5.354^{\circ}$) in the energy range from 500~MeV to 100~GeV. The low energy threshold of 500~MeV was chosen to minimize the possible contamination of any nearby point sources. Based on the on-orbit performance\footnote{\texttt{http://www.slac.stanford.edu/exp/glast/groups/canda/\\lat\_Performance.htm}} of \fermi, the 95\% containment angle (i.e., the point spread function) for all normal incidence events at $E>500$~MeV is $\sim3.5^{\circ}$. The only known LAT source within 3.5$^{\circ}$ of \src\ is the blazar 2FGL~J0426.6+0509c that is 1.63$^{\circ}$ away.

To model the ROI, we included the standard diffuse templates\footnote{\texttt{gll\_iem\_v05\_rev1.fit} and \texttt{iso\_source\_v05.txt}, which are taken from \texttt{http://fermi.gsfc.nasa.gov/ssc/data/access/lat/\\BackgroundModels.html}} to model the Galactic and isotropic background, as well as all sources in the 2FGL catalog \citep{Nolan12} within 10$^{\circ}$ of \src. Note that \src\ is not included in the 2FGL catalog.
The spectral shapes of the latter were fixed to their 2FGL values but the normalizations were set free. 2FGL sources within the annulus of 10$^{\circ}$--15$^{\circ}$ were also included, but all of their parameters were fixed to their 2FGL values. By performing \texttt{gtlike} for the six-year accumulated data, we first found a test statistic (see \citet{Mattox96}), ${\rm TS}=47.6$ for \src\ with power-law index $\Gamma=2.95 \pm 0.22$ and 0.5--100 GeV flux $\left( 1.8 \pm0.3\right)\times10^{-9}$ photons cm$^{-2}$ s$^{-1}$, corresponding to 0.1--100 GeV flux of $\left( 4.2 \pm0.7 \right)\times10^{-8}$ photons cm$^{-2}$ s$^{-1}$ when extrapolated down to 100~MeV by assuming $\Gamma=2.95$. The large index compared to the average index $\Gamma\sim2.4$ of LAT-detected radio galaxies \citep{Abdo10MAGN, Inoue11} might be due to the selection of the higher low-energy threshold of 500~MeV instead of 100~MeV.  We determined the LAT source position for \src\ using \texttt{gtfindsrc} and obtained R.A.=68.287$^{\circ}$ and Decl.=5.381$^{\circ}$ (68\% error circle radius 0.088$^{\circ}$) which is only 0.028$^{\circ}$ away from the radio position. A new LAT source was found with ${\rm TS}=103$ at ${\rm R.A.}=68.7^{\circ}$, ${\rm Decl.}=9.4^{\circ}$, $\sim 4.1^{\circ}$ away from \src.
This source was included in the source model during the LAT analysis, in addition to the 2FGL sources.
Light curves and spectral data points were computed by dividing the data set in time bins (30 and 5 days) or energy bins (3 logarithmic bins from 0.5 to 100 GeV) and fixing the parameters of all background sources (except for the normalizations of two nearby sources which might be variable) and leaving the normalization of \src\ free. In both cases, the spectral index of \src\ and the two nearby background sources were fixed to the value obtained from the full six-year analysis. A gtlike analysis was then performed in each bin to determine the normalization of the \src\ power law. 90\% upper limits were computed in any bin where the TS was less than 9.

\subsection{Submillimeter}
\label{sec-submm}

The 230~GHz (1.3 mm) light curve for \src\ was obtained using the SMA near the summit of Mauna Kea (Hawaii). \src\ is included in an ongoing SMA monitoring program to determine the flux densities of compact extragalactic radio sources that can be used as millimeter and submillimeter wavelength calibrators \citep{Gurwell07}. 
The observations were obtained in the `compact' array configuration with typical 3$\arcsec$ resolution.
Data from this program are updated regularly and are available on the SMA website\footnote{\texttt{http://sma1.sma.hawaii.edu/callist/callist.html}}.

\subsection{Radio}
\label{sec-radio}

We retrieved all available 43~GHz VLBA data from the Boston University Monitoring Program from 2012 January to 2013 December and analyzed the calibrated $(u,v)$ data using \texttt{DIFMAP}. The intensity of the core was measured in images convolved with a common Gaussian beam of $0.45\times0.20$~mas to avoid resolution effects. We found a clear enhancement at epoch ${\rm MJD}=56307$, which was followed by an ejection of a new jet component and decreasing core intensity in subsequent epochs. Using the visibility-based model-fitting program in \texttt{DIFMAP} to model the intensity structures, we found the new component was unresolved from the core on ${\rm MJD}=56307$, but was clearly identified on the subsequent two epochs (${\rm MJD}=56349$, 56398). This suggests its epoch of zero separation was around ${\rm MJD}=56307$ or slightly later.

\section{$\gamma$-ray and Radio Correlation}
\label{sec-results}

\begin{figure}[!th]
\begin{center}
\includegraphics[width=0.5\textwidth]{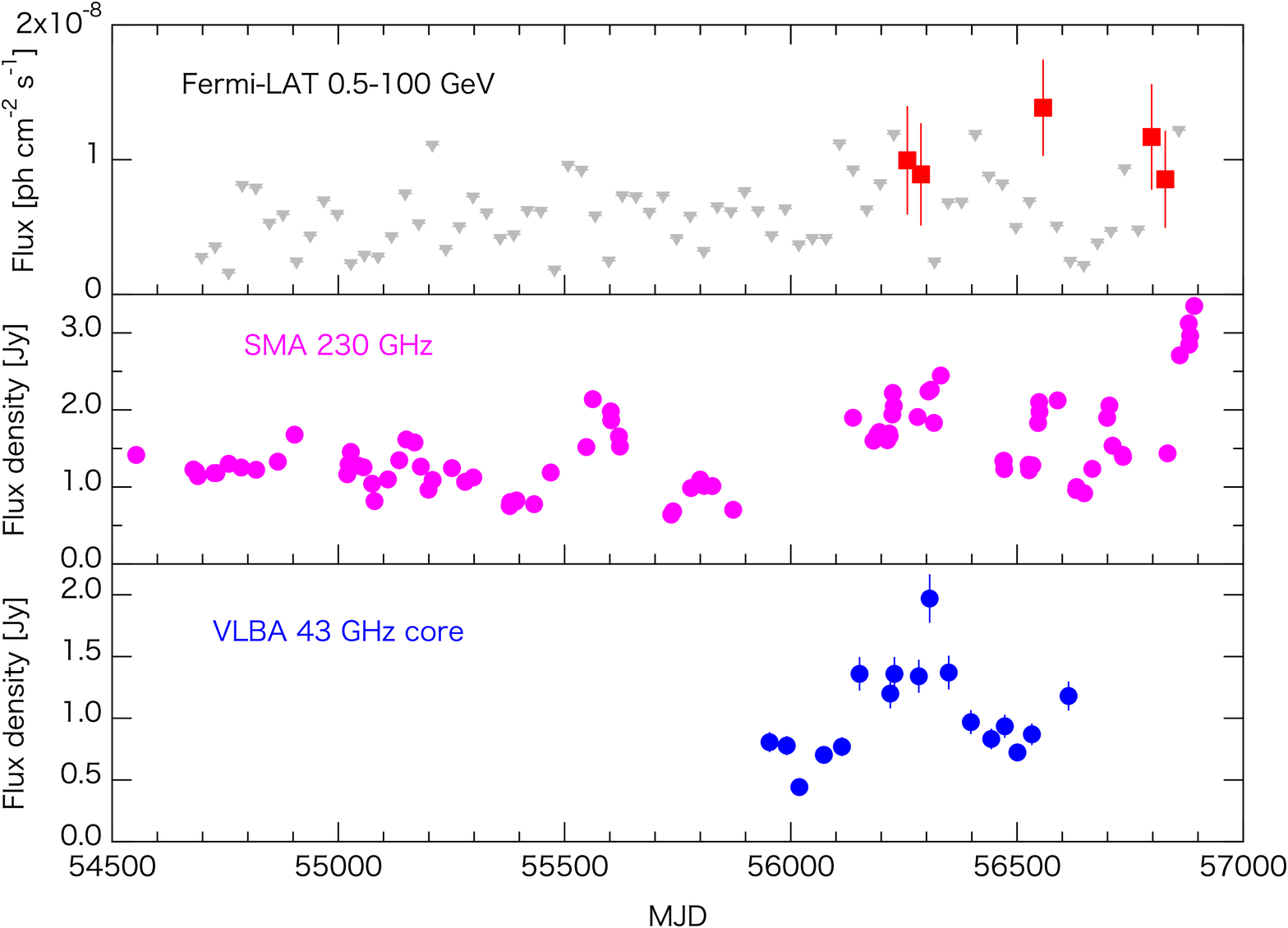}
\caption{High-energy $\gamma$-ray (30-day-bin), 230~GHz SMA, and 43~GHz VLBA core light curves of \src\ from MJD 54500 to 57000 (corresponding to 2008 February 4 and 2014 December 9, respectively). In the top panel, \src\ is measured with ${\rm TS}>9$ for only five time bins (red squares) and 90\% confidence level flux upper limits are shown when ${\rm TS}<9$ (gray triangles).}
\label{fig:lc}
\end{center}
\end{figure}

\begin{figure}[!th]
\begin{center}
\includegraphics[width=0.5\textwidth]{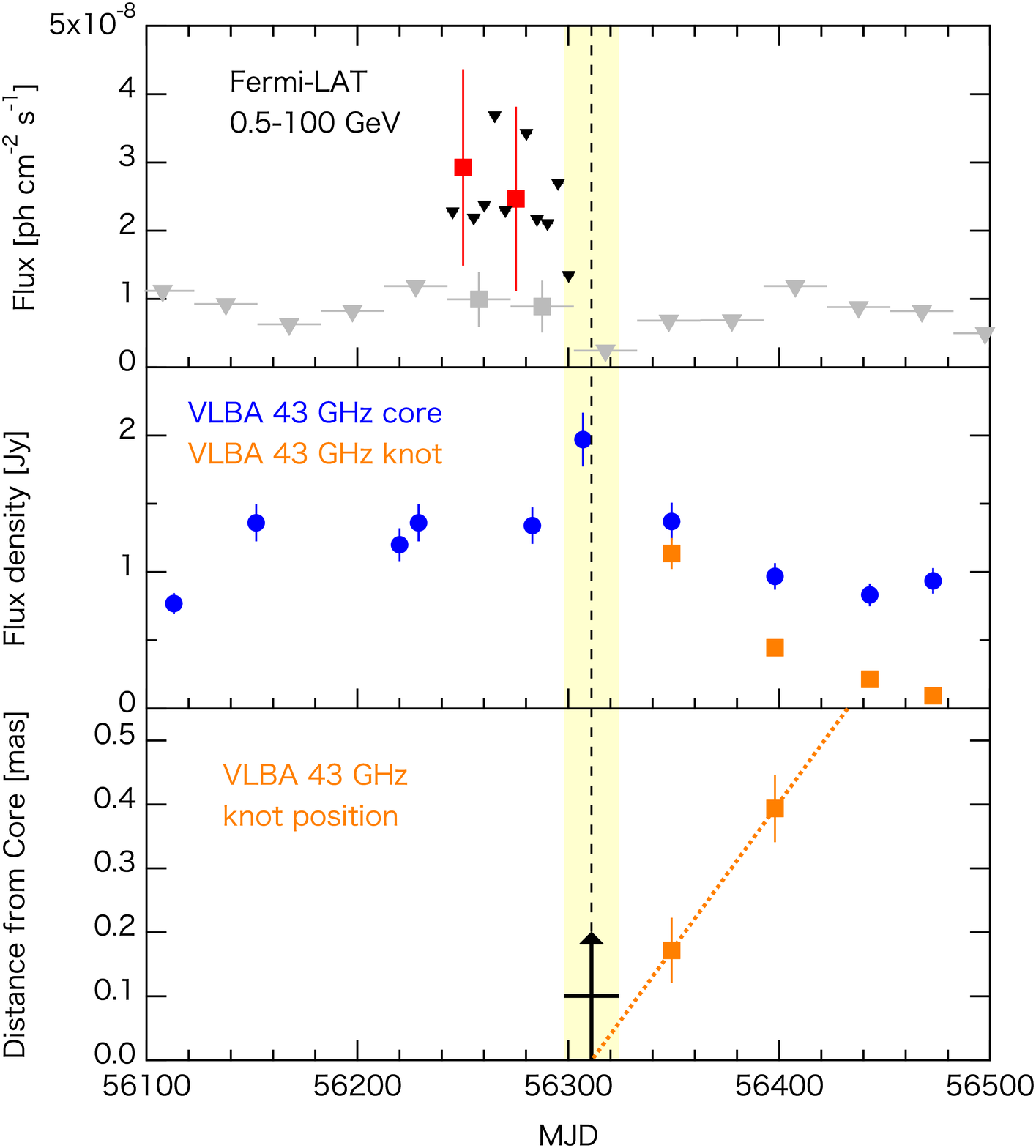}
\caption{{\it Top}: \fermi\ 0.5--100~GeV fluxes of \src\ calculated in 5-day bins with 90\% confidence level flux upper limit shown by the black triangle when ${\rm TS}<4$. Gray squares and triangles are the 30-day binned \fermi\ data shown in the {\it Top} panel of Figure~\ref{fig:lc}. {\it Middle}: Light curves of VLBA 43~GHz core (blue circles) and jet knot (orange squares). {\it Bottom}: Angular separation of the jet knot from the core as a function of time. The orange dashed line shows the best-fit for the two data points. The black arrow and black dashed line indicate the time when the knot passed through the core, and the yellow hatched area is the associated error (MJD 56311$\pm$13).}
\label{fig:lc_zoom}
\end{center}
\end{figure}

Fig.~\ref{fig:lc} shows the 30-day-bin 0.5--100~GeV flux of \src, together with the 1-day-bin 230~GHz and VLBA 43~GHz core flux light curves. Here we assume that the SMA fluxes are dominated by the core. \fermi\ measured a ${\rm TS}>9$ from \src\ during the two most recent years, while the 230~GHz light curve indicates an enhanced activity of the source in the second half of the analyzed six years. In Fig.~\ref{fig:lc_zoom}, we display enlarged views of the \fermi\ 0.5--100~GeV and VLBA 43~GHz light curves between MJD 56100 and 56500. In the 5-day binned \fermi\ light curve (see {\it top} panel), the largest values of ${\rm TS}=7.4$ and 7.5 are found  in two bins, MJD 56247.7--56252.7 and 56272.7--56277.7, respectively.

\begin{figure}[!t]
\begin{center}
\includegraphics[width=0.3\textwidth]{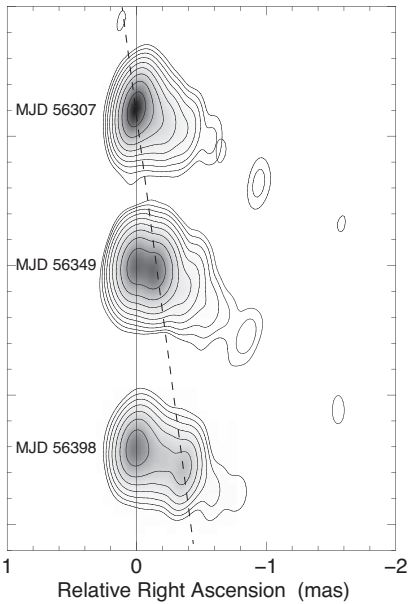}
\caption{Sequence of 43~GHz VLBA images of \src\ at epochs MJD 56307, 56349, and 56398. The vertical axis shows the epoch in linear scale. The dashed line indicates the trajectory of the ejected knot.}
\label{fig:vlba}
\end{center}
\end{figure}

Fig.~\ref{fig:vlba} shows the 43~GHz VLBA images at three epochs obtained after the \fermi\ detection of MeV/GeV $\gamma$ rays. A bright radio knot clearly emerged from the core and propagated down the jet. The 43~GHz light curves of the core and the knot are shown in Fig.~\ref{fig:lc_zoom} along with the separation of the knot from the core as a function of time. We derived an apparent velocity of the knot as $3.5\pm0.9 c$ by fitting the first and second data points, because the low fluxes in the third and fourth points did not allow us to precisely determine the knot position. The apparent speed derived here is similar to past VLBA measurement of $4.0\pm0.2c$ estimated from moving components with well-determined motions \citep[][Table~5 therein]{Jorstad05}. Extrapolating the motion of the ejected knot linearly, we infer that it passed through the core at MJD 56311$\pm$13, after the MeV/GeV flare occurred (see Fig.~\ref{fig:lc_zoom}). During the passage, the 43~GHz core flux reached a maximum. Detailed VLBA analysis results will be reported elsewhere (Doi et al. in prep).

\section{Discussion}
\label{sec-dis}
Although we can not rule out coincidence, the fact that the $\gamma$-ray flares on MJD 56247.7-56252.7 and 56272.7-56277.7 were followed in close temporal proximity by the radio core brightening and superluminal radio ejection on MJD $56311\pm13$ leads us to assume that they are physically related. In this regard, we also note that the 43~GHz core flux again increased after the third $\gamma$-ray flare occurred around MJD~56560 (see the lower panel in Fig.~\ref{fig:lc}). This implies that a knot, responsible for the second $\gamma$-ray flare, was passing through the core and hence the 43~GHz core flux increased around MJD~56600, further supporting the physical connection between the $\gamma$-ray and radio phenomena. The distance from the central BH to the VLBA core was estimated to be $\sim 0.5$~pc by using the time lag between the X-ray dip and radio jet ejection \citep{Chatterjee09}. The apparent speed of the knot of $3.5\pm0.9 c$ derived here and the time lags of $\sim60$ days and $\sim35$ days between the $\gamma$-ray and VLBA core brightenings allow us to infer the position of the $\gamma$-ray emission region under the assumption that these two phenomena are physically related. The knot moves a projected distance of $\sim0.18$ and $\sim0.11$~pc in the $\sim60$ and $\sim35$ days, respectively, while \citet{Chatterjee09} derived a projected distance of 0.22 pc from the central BH to the 43~GHz VLBA core. Given the angle of $\sim20.5^{\circ}$ between the jet axis and our line of sight \citep{Jorstad05}, we found that the $\gamma$-ray emission region is located at $\sim0.1$ and $\sim0.3$ pc from the central BH. Here we assumed that the knot velocity is constant and if the acceleration of the knot takes place between the BH and VLBA core, as speculated by e.g., \citet{Marscher08}, the distance from the BH becomes larger, but still on the order of sub-pc. 

It is difficult to precisely determine a source variability timescale from the \fermi\ data due to the relatively low flux. But since the \fermi\ sporadic detections are dominated by several 5-day bins (see top panel of Fig.~\ref{fig:lc_zoom}), we assume here that the variability timescale is 5--10 days. The derived distance between the BH and $\gamma$-ray emission region, together with the assumed variability timescale of 5--10 days and previously-measured beaming factor $\delta=2.4$ \citep{Jorstad05}, allow us to estimate the jet opening angle. The radius of the emission region is estimated as $R\lesssim c \delta t_{\rm var} =4.7 \times 10^{16} \left(\delta/2.4\right) \left(t_{\rm var}/7.5 \,{\rm day}\right)$ cm. By assuming a simple conical shape for the jet structure, the jet half opening angle ($\theta_{\rm jet}$) is estimated as $\theta_{\rm jet} \sim R/r \sim 4.5^{\circ}$, where the distance from the BH to the $\gamma$-ray emission region, $r=0.2$~pc is assumed by taking the average of the two distances derived above. This is consistent with the past estimate of the jet half opening angle of $3.8\pm1.3$ degree based on sequential VLBA observations \citep[see Table~11 of][]{Jorstad05}. The bulk Lorentz factor $\Gamma\sim4$ is estimated from the apparent speed of the knot of $3.5c$ and viewing angle of 20.5$^{\circ}$. Hence, combined with $\theta_{\rm jet}\sim4.5^{\circ}$ (or 0.078~rad) derived here, we obtain $\Gamma \theta_{\rm jet}\sim0.3$, which is similar to the recently claimed result of $\Gamma \theta_{\rm jet}\approx 0.2$ from the MOJAVE study \citep{Clausen13}. 

\begin{figure}[!th] 
\begin{center}
\includegraphics[width=0.5\textwidth]{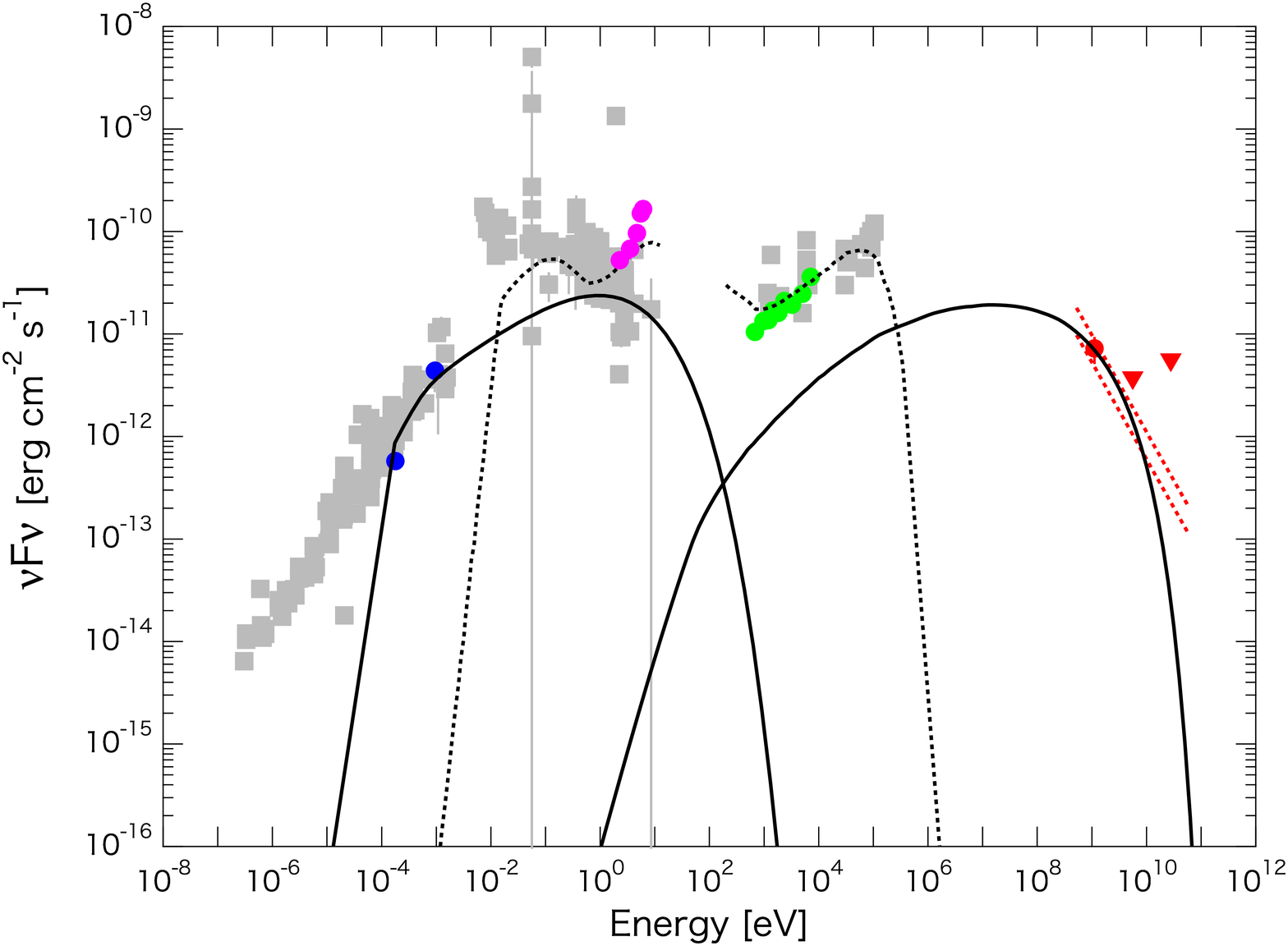}
\caption{\fermi\ spectrum of \src\ during the 60-day flaring state in MeV/GeV band (MJD 56242.7 to 56302.7; red circle and triangles are fluxes and 95\% confidence upper limits) together with historical data taken from NED database (gray squares). The red dashed lines indicate the 1$\sigma$ normalization error limit and here we fixed the power-law index to 2.95 (derived from the six-year data). Blue circles represent contemporaneous 43~GHz VLBA and 230~GHz core fluxes measured on MJD 56283 and 56280, respectively. Magenta and green circles are the {\it Swift}/UVOT and XRT fluxes observed on MJD~56276. The synchrotron + SSC emission model is overlaid with the solid lines while the dashed line indicate the template of accretion-related Seyfert-type emission, taken from \citet{Koratkar99}.
}
\label{fig:sed}
\end{center}
\end{figure}

Fig.~\ref{fig:sed} shows the \fermi\ spectrum of \src\ during the 60 day (MJD 56242.7--56302.7) $\gamma$-ray flaring period. The temporally closest measurements of the 43~GHz VLBA core and 230~GHz fluxes measured on MJD 56283 and 56280.2 are also plotted, together with the \swift/UVOT and XRT fluxes on MJD~56276. Note that this is the first truly simultaneous broadband spectrum for this source. The disk emission of \src\ is known to be bright \citep[e.g.,][]{Kataoka07} and the accretion rate is estimated to be as large as $\sim10$\% of its Eddington luminosity \citep[e.g.,][]{Chatterjee09}. This implies a possible contribution of an external photon field as targets for the inverse-Comptonization process. In the following, we constrain the possible target photon field through energetic arguments. The ratio between the external-radiation Compton (ERC) luminosity to SSC luminosity is $L_{\rm ERC}/L_{\rm SSC}=(\delta/\Gamma)^2(u'_{\rm ext}/u'_{\rm syn})$, where $u'_{\rm ext}$ and $u'_{\rm syn}$ are the external and synchrotron photon energy density in the jet comoving frame, respectively \citep{Stawarz03}. Here we consider only the circumnuclear hot dust region (HDR) as an external photon field (namely, $u'_{\rm ext}\sim u'_{\rm HDR}$). Because the location of the MeV/GeV $\gamma$-ray emission region (0.1--0.3~pc, see \S\ref{sec-results}) is well beyond the broad line region (BLR) radius of 0.019--0.024~pc \citep[determined from reverberation mapping by][]{Pozo14}, we can neglect the possible contribution of BLR photons.

The energy density of the HDR emission in the jet comoving frame is $u'_{\rm HDR}\approx\Gamma^2L_{\rm HDR}/4\pi r_{\rm HDR}^2c$. Here, the HDR radius is estimated as $r_{\rm HDR}\simeq 4 \left( L_{\rm disk}/10^{46} \ {\rm erg/s} \right)^{1/2} \ {\rm pc}\sim0.6$~pc, using the disk flux of $\sim10^{-10}$ erg cm$^{-2}$ s$^{-1}$ (and hence $L_{\rm disk}\sim2 \times 10^{44}$ erg s$^{-1}$) in the UV band (see Fig.~\ref{fig:sed} and below). We also obtain the HDR luminosity as $L_{\rm HDR}\sim\xi_{\rm HDR} L_{\rm disk}=2 \times 10^{43} \left( \xi_{\rm HDR}/0.1 \right)$ erg s$^{-1}$, where $\xi_{\rm HDR}$ is the fraction of the disk luminosity reprocessed in the inner region of the hot dust torus ($\sim10$\% is usually assumed, \citet{Sikora09}). The synchrotron photon energy density in the jet comoving frame is $u'_{\rm syn}=L_{\rm syn}/4\pi \delta^4 R^2c$ and the peak flux is roughly $\sim 10^{-11}$ erg cm$^{-2}$ s$^{-1}$ (see Fig.~\ref{fig:sed} and below), hence $L_{\rm syn} \sim 2 \times 10^{43}$ erg s$^{-1}$. Combining these, we obtained
\begin{eqnarray}
\frac{L_{\rm ERC}}{L_{\rm SSC}} & \approx & c^2\delta^8t_{\rm var}^2r_{\rm HDR}^{-2} L_{\rm HDR} L^{-1}_{\rm syn} \nonumber \\
& \sim & 0.1 \left(\frac{\delta}{2.4} \right)^8 \left( \frac{t_{\rm var}}{7.5 \ {\rm day}} \right)^2 \left(\frac{r_{\rm HDR}}{0.6 \ {\rm pc}} \right)^{-2} \left(\frac{\xi_{\rm HDR}}{0.1} \right) \nonumber \\ 
& \times & \left( \frac{L_{\rm disk}}{2\times10^{44} \ {\rm erg/s}} \right) \left( \frac{L_{\rm syn}}{2\times10^{43}\ {\rm erg/s}} \right)^{-1}. \nonumber
\end{eqnarray}
Therefore, despite the bright disk and dusty torus, the SSC scenario may be favored over the ERC as the MeV/GeV $\gamma$-ray emission mechanism. 
Note the strong dependence of this estimate on the Doppler factor where we took the average $\delta=2.4\pm0.6$ derived by \citet{Jorstad05} based on a VLBA kinematic study over 3 years. If we take the upper end of their average, $\delta=3.0$ instead of $\delta=2.4$ assumed above, the luminosity ratio increases to $L_{\rm ERC}/L_{\rm SSC} \sim 0.6$, which is closer to unity. 

To derive the physical parameters at the emission site, we performed SED modeling for the 43~GHz VLBA, 230~GHz SMA, and \fermi\ data points. 
We assumed one-zone synchrotron and SSC emissions \citep{Finke08} by relativistic electrons of a single power-law distribution with exponential cutoff, $dN/d\gamma=K\gamma^{-s}\exp(-\gamma/\gamma_{\rm cut})$, where $K$ is the normalization, $s$ is the electron spectral index, and $\gamma_{\rm cut}$ is the cutoff electron Lorentz factor. We first fixed the following parameters, $\delta=2.4$, $\Gamma=4.0$, and $R=4.7 \times 10^{16}$~cm (corresponding to $t_{\rm  var}=7.5$ days). Then we searched for a set of parameters well representing the simultaneous multi-wavelength data; minimum and maximum electron Lorentz factors $\gamma_{\rm min}$ and $\gamma_{\rm max}$, $\gamma_{\rm cut}$, $K$, $s$, and comoving magnetic field $B$. As a result, we obtained $\left( \gamma_{\rm min}, \gamma_{\rm cut}, \gamma_{\rm max} \right) = \left( 10^2, 7.5\times10^3, 10^6 \right)$, $K=8.0\times10^{54}$ electrons, $s=2.2$, and $B=1.0$ Gauss.
The synchrotron curve is largely unconstrained due to the sparse coverage with only 43 and 230~GHz measurements. However, \citet{Wolk10} reported a significant jet contribution to the mid-IR flux, in agreement with the modeling presented here.
These parameters derived from SED modeling lead to the jet powers in electrons ($L_{\rm e}$) and magnetic field ($L_{\rm B}$) of $5.2\times10^{43}$ erg s$^{-1}$ and $1.3\times10^{44}$ erg s$^{-1}$, respectively. Hence Poynting flux slightly dominates, $U^{\prime}_e/U^{\prime}_B \sim 0.4$, where $U^{\prime}_e$ and $U^{\prime}_B$ are comoving electron and magnetic energy density, assuming no dynamically relevant protons in the jet. 
The total radiated power ($L_{\rm rad}$) based on the SED modeling was calculated as $L_{\rm rad}=(4/3)(\Gamma^2/\delta^4)(L_{\rm syn}+L_{\rm SSC})=4.9 \times 10^{44}$ erg s$^{-1}$ by assuming $\delta=2.4$ and $\Gamma=4.0$. This is larger than the summed $L_{\rm e} + L_{\rm B}$ and hence another source of power is required.
A possible solution is to assume that the jet carries a cold proton component which has larger power than $L_{\rm rad}$ \citep[e.g.,][]{Sikora00,Ghisellini14}.
Indeed, there are some arguments justifying the need for a proton component \citep[][and references therein]{Celotti93,Ghisellini10}. If we assume that the total radiated power is about 10\% of the jet power, as was found for AGN and Gamma-Ray Burst jets \citep{Nemmen12}, we find the jet power is $L_{\rm jet} \sim 5 \times 10^{45}$ erg s$^{-1}$. 
On the other hand, the accretion luminosity, defined as the total power of accreting plasma, is estimated as $L_{\rm acc} \simeq 2 \times 10^{45}$ erg s$^{-1}$ from the disk luminosity of $2 \times 10^{44}$ erg s$^{-1}$ and assuming a 10\% radiative efficiency of the disk. Hence, we obtain $L_{\rm jet} \gtrsim L_{\rm acc}$, indicating that the jet launching and accelerating process is extremely efficient \citep[see also e.g.,][]{Tanaka11,Saito13,Ghisellini14}. Simultaneous optical/UV and X-ray fluxes are above the jet component (see Fig.~\ref{fig:sed}) and therefore originate from the disk even during the high $\gamma$-ray state. Indeed, the optical/UV/X-ray spectral shape is roughly consistent with accretion-related Seyfert-type emission \citep{Koratkar99}.

\acknowledgments

The \textit{Fermi}-LAT Collaboration acknowledges support for LAT development, operation and data analysis from NASA and DOE (United States), CEA/Irfu and IN2P3/CNRS (France), ASI and INFN (Italy), MEXT, KEK, and JAXA (Japan), and the K.A.~Wallenberg Foundation, the Swedish Research Council and the National Space Board (Sweden). Science analysis support in the operations phase from INAF (Italy) and CNES (France) is also gratefully acknowledged.

This study makes use of 43~GHz VLBA data from the VLBA-BU Blazar Monitoring Program (VLBA-BU-BLAZAR; http://www.bu.edu/blazars/VLBAproject.html), funded by NASA through the \fermi\ Guest Investigator Program. The VLBA is an instrument of the National Radio Astronomy Observatory, a facility of the National Science Foundation operated by Associated Universities, Inc.

The Submillimeter Array is a joint project between the Smithsonian Astrophysical Observatory and the Academia Sinica Institute of Astronomy and Astrophysics and is funded by the Smithsonian Institution and the Academia Sinica.

C.C.C. was supported at NRL by NASA DPR S-15633-Y. {\L}.S. was supported by Polish NSC grant DEC-2012/04/A/ST9/00083.

\bibliographystyle{apj}

\end{document}